\documentclass[onecolumn,11pt]{article}
\usepackage[top=1in, bottom=1in, left=1in, right=1in]{geometry}
\usepackage{amsmath,amsfonts,amscd,amssymb}
\usepackage{graphicx}
\usepackage{epstopdf}
\usepackage{overpic}
\usepackage{cancel}
\usepackage{rotating}
\usepackage{url}
\usepackage{caption}
\usepackage{color}
\usepackage{rotating}
\usepackage{multirow}
\usepackage{wrapfig}
\usepackage{mathtools}
\usepackage{subeqnarray}
\usepackage{setspace}
\usepackage{palatino} 
\usepackage{subcaption}
\usepackage[numbers,sort&compress]{natbib}

\usepackage[bottom,flushmargin,hang,multiple]{footmisc}
\usepackage{lipsum}

\definecolor{header1}{cmyk}{0,0,0,1}

\usepackage{dcolumn}
\usepackage{bm}
\usepackage{subcaption}
\usepackage{hyperref}
\usepackage{xcolor}

\title{\LARGE{\vspace{-.55in}\textbf{Advanced Modeling for the HIT-SI Experiment}}\vspace{-.175in}}

\author{\normalsize{A. A. Kaptanoglu$^{1}$, T. E. Benedett$^2$, K. D. Morgan$^2$, C. J. Hansen$^{2,3}$, T. R. Jarboe$^2$}\\
\footnotesize{$^1$ Department of Physics, University of Washington, Seattle, WA 98195, United States}\\
\footnotesize{$^2$ Department of Aeronautics and Astronautics, University of Washington, Seattle, WA 98195, United States}\\
\footnotesize{$^3$ Department of Applied Physics and Applied Mathematics, Columbia University, New York, NY 10027, United States}
}
\date{}
\begin{document}
\maketitle
\begin{abstract}
A two-temperature magnetohydrodynamic (MHD) model, which evolves the electron and ion temperatures separately, is implemented in the PSI-Tet code and used to model plasma dynamics in the HIT-SI experiment. When compared with single-temperature Hall-MHD, the two-temperature Hall-MHD model demonstrates improved qualitative agreement with experimental measurements, including: far-infrared interferometry, ion Doppler spectroscopy, Thomson scattering, and magnetic probe measurements. 
The two-temperature model is utilized for HIT-SI simulations in both the PSI-Tet and NIMROD codes at a number of different injector frequencies in the $14.5-68.5$ kHz range. At all frequencies the two-temperature models result in increased toroidal current, lower chord-averaged density, and symmetrization of the current centroid, relative to single-temperature simulations. Both codes produce higher average temperatures and toroidal currents as the injector frequency is increased. Power balance and heat fluxes to the wall are calculated for the two-temperature PSI-Tet model and indicate considerable viscous and compressive heating, particularly at high injector frequency. 
Parameter scans are also presented for the artificial diffusivity, and Dirichlet wall temperature and density. Artificial diffusivity and the density boundary condition both significantly modify the plasma density profiles, leading to larger average temperatures, higher toroidal current, and increased relative density fluctuations at low diffusivity and low wall density. High power, low density simulations at 14.5 kHz achieve sufficiently high gain ($G\approx 5$) to generate significant volumes of closed flux lasting 1-2 injector periods. 
\\


\end{abstract}

\section{\label{sec:level1}Introduction}
Space and laboratory plasmas are almost universally multi-scale and strongly nonlinear. This motivates the use of simulations to improve understanding, but necessarily requires complex numerical models. Efficiency and model validation, balancing complexity and computational speed, is essential for simulating plasmas from solar wind dynamics to the interiors of fusion experiments. Magnetohydrodynamic (MHD) codes have now been used in the plasma physics community for over 50 years, but the complexity and inherent scale separations result in many numerical codes which are specialized for the efficient computation of a specific type of experiment or plasma system. For instance, a number of codes are optimized for solving in axial and toroidal geometry~\cite{sovinec2004nonlinear,schnack1987semi,park1999plasma}. As performance and efficiency have improved, additional physical effects have been added to the basic single-fluid MHD model, such as: two-fluid (Hall) effects, finite Larmor radius (FLR) effects, evolution of a neutral fluid, and improved closures. The importance of these terms for HIT-SI simulations is chosen from theoretical considerations~\cite{jarboe2012imposed} and the results of experimental validation~\cite{izzo2003numerical,akcay2013extended,morgan2017validation}. Investigating the differences between plasma systems with a collection of models and geometries of varying complexities facilitates new understanding of the plasma dynamics, as well as validation of the numerical codes. The work presented here is part of an effort to investigate the complexity of the physical models, geometry, and boundary conditions required to accurately simulate the HIT-SI experiment, while retaining high efficiency. 

Previous work on HIT-SI simulations started with a resistive single fluid model~\cite{izzo2003numerical} followed by a model including Hall terms to capture some of the effects expected with the full two-fluid system of separate ions and electrons~\cite{akcay2013extended}; these models are typically referred to as resistive-MHD and Hall-MHD respectively. Two-fluid effects through the Hall terms are expected to be important for spatial scales between the ion and electron inertial scales~\cite{freidberg2014ideal}. In HIT-SI the ion inertial scale $d_i \approx 8$ cm is comparable with experimental length scales, such as the diameter of an injector mouth $d_\text{inj} \approx 14$ cm, and the characteristic magnetic scales of the spheromak $\lambda_\text{sph}^{-1} \approx 10$ cm, and of the injector $\lambda_\text{inj}^{-1} \approx 5$ cm~\cite{jarboe1994review,bellan2000spheromaks}. 
Numerical models with the Hall terms significantly improve agreement with HIT-SI measurements over the resistive MHD models~\cite{akcay2013extended}, and validate well with many bulk measurements of the experiment~\cite{morgan2017validation,TomB}. However, from measurements of HIT-SI~\cite{hossack2015study} (and a newer device called HIT-SI3)~\cite{everson2019hit} and theory~\cite{jarboe1994review,jarboe2006spheromak}, ion temperatures are expected to be significantly higher than electron temperatures. Distinguishing between the ion and electron temperatures provides additional validation with experiment over the single temperature model, and leads to new insights into the energy flow and dynamical processes in the system. 

\subsection{The HIT-SI experiment}
\begin{figure*}[!tp]
    \centering
    \begin{subfigure}{0.45\textwidth}
    \includegraphics[width=0.9\textwidth]{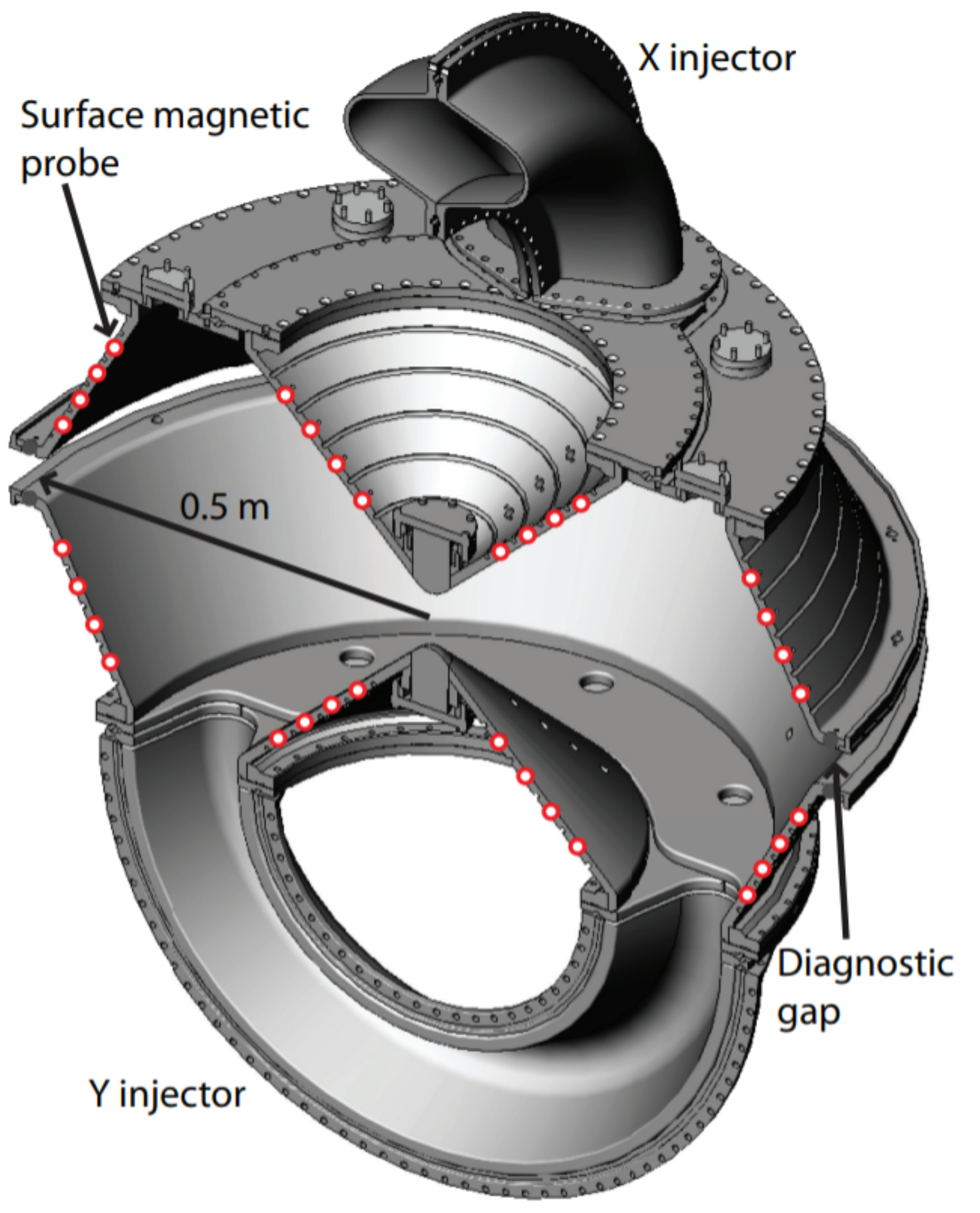}
    \end{subfigure}
    \begin{subfigure}{0.45\textwidth}
    \includegraphics[width=0.9\textwidth]{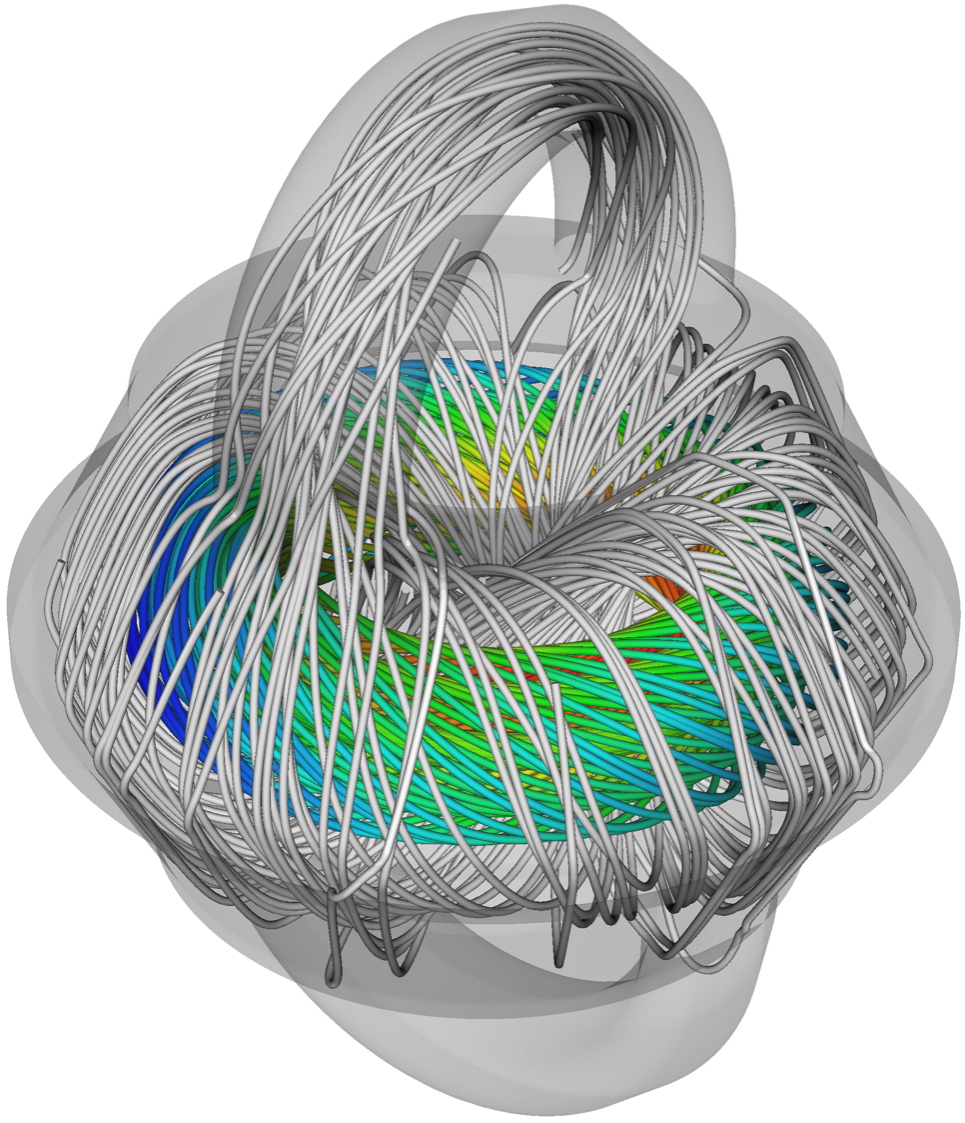}
    \end{subfigure}    
    \caption{Left: A cross section of the device shows the toroidal structure, the two magnetic helicity injectors, and the surface probe locations. Reproduced from Wrobel et al., Relaxation-time measurement via a time-dependent helicity balance model, Physics of Plasmas, 20(1):012503, 2013, with the permission of AIP Publishing. Right: Representative equilibrium during sustainment with an injector shows an axisymmetric spheromak (rainbow) surrounded by field lines tied to the injector (gray).}
    \label{fig:HIT-SI}
\end{figure*}

HIT-SI was a laboratory device that formed and sustained spheromak plasmas for the study of plasma self-organization and steady inductive helicity injection (SIHI)~\cite{jarboe2006spheromak}. 
It consisted of an axisymmetric flux conserver and two inductive injectors (called the X-injector and the Y-injector) mounted on each end as shown in Fig. \ref{fig:HIT-SI}. Voltage and flux circuits are wrapped around the outside of the injectors but are not pictured here. The voltage and axial flux of each injector are oscillated in phase at a frequency $f_\text{inj}$ with values between $14.5-68.5$ kHz. The two injectors are spatially and temporally 90$^o$ out of phase and are purely inductive. The power and magnetic helicity injected by the two injectors is approximately constant during the discharge and is slowly varying compared with $f_\text{inj}$. An experimental discharge begins with an initial formation period, followed by the nonlinear self-organization of the spheromak plasma, which is subsequently sustained by the injectors for the remainder of the discharge. A description of the equilibrium profile and postulated current drive mechanism during sustainment can be found in the references on imposed dynamo current drive (IDCD)~\cite{jarboe2012imposed,jarboe2014proof}. Detailed descriptions of the experiment can be found in Jarboe et. al.~\cite{jarboe2006spheromak} and Wrobel~\cite{wrobel2011study}. The upgraded HIT-SI3 device is dynamically similar, but features three inductive injectors mounted solely on the top of the device. 

\subsection{Contributions of this work}
A two-temperature Hall-MHD model is implemented in the PSI-Tet extended MHD code. The PSI-Tet and NIMROD models are described in Section \ref{sec:modeling} and are compared with their single-temperature counterparts. In Section \ref{sec:freqscan}, we extend the scans of the injector frequency performed in previous works~\cite{morgan2018finite,TomB} and find a number of interesting and important changes in the plasma dynamics, including: significantly hotter ions than electrons at high frequency operation, new insights into the dynamical power flows, super-linear injector impedance scaling with the injector frequency, and improved understanding of the structure of spheromak formation. When compared to the single-temperature models, the two-temperature models robustly predict larger (or similar magnitude) toroidal currents, increased volume-averaged ion temperatures, lower chord-averaged densities, and an outward radial shift (this is only true for PSI-Tet) and vertical symmetrization of the current centroid. Moreover, we discover that modeling the injector geometry in PSI-Tet appears to be important for reproducing the experimentally observed asymmetry between injector waveforms and corresponding vertical shift of the current centroid; the NIMROD single temperature model, which does not model the injector geometry, observes a slight vertical shift but this shift is not present with the two-temperature NIMROD model. 

In Section \ref{sec:paramscan}, parameter space scans are performed for the artificial diffusivity and the density and temperature Dirichlet boundary conditions. The wall temperature scan indicates an inward shift of the current centroid and reduced volume-averaged $\beta$, consistent with increased thermal pressure near the wall. Scans in the density and artificial diffusivity indicate operating in a low density regime leads to a hotter plasma with higher toroidal currents and larger relative density fluctuations, consistent with experimental results and theoretical work on IDCD~\cite{jarboe2012imposed,jarboe2014proof}. Low frequency, low density, high power simulations illustrate evidence of a transition to a high performance regime, with closed flux events lasting $50-100$ $\mu$s (or 1-2 injector periods, $\tau_\text{inj}$) at spheromak gains of $G \approx 5$. In Section \ref{sec:conclusion}, we conclude with a number of suggestions for future work to improve the modeling and validation capabilities of these codes. The code used in this work can be found at \url{https://github.com/akaptano/Two-Temperature-HITSI-Analysis}.

\section{\label{sec:modeling}Details of the MHD model}
Numerical modeling of HIT-SI with MHD is a critically enabling tool for physical understanding of this dynamic system and design of future current drive experiments based on SIHI. We utilize two MHD codes, NIMROD~\cite{sovinec2004nonlinear} and PSI-Tet~\cite{hansen2015simulation,hansen2014mhd}, which differ primarily in their spatial discretizations. 

\subsection{MHD model}
The first-principles model must balance efficiency and complexity, and this balance will vary depending on the physics required to resolve the dynamics of interest. 
The two-temperature Hall-MHD model chosen for these simulations was informed by previous work~\cite{akcay2013extended,morgan2018finite} and theoretical considerations~\cite{jarboe2012imposed,jarboe2014proof}.
This model is defined by the following evolution equations for the plasma density $n$, fluid velocity $\bm{u}$, ion temperature $T_i$, electron temperature $T_e$, and magnetic field $\bm{B}$,
\begin{equation} \label{eq:xmhd_system_1}
\notag
\frac{\partial n}{\partial t} = - \nabla \cdot \left( n \bm{u} \right) + D \nabla^2 n,
\end{equation}
\begin{equation} \label{eq:xmhd_system_2}
\notag
\frac{\partial \bm{u}}{\partial t} = - \bm{u} \cdot \nabla \bm{u} + \frac{1}{m_i n} \left[ \bm{J} \times \bm{B} - \nabla \left( n(T_i+\textcolor{blue}{T_e}) \right)
- \nabla \cdot \bm{\Pi} \right],
\end{equation}
\begin{equation} \label{eq:xmhd_system_3}
\notag
\frac{\partial T_i}{\partial t} = - \bm{u} \cdot \nabla T_i + \left( \gamma-1 \right) \left[ -T_i \nabla \cdot \bm{u}
- \frac{1}{n} \left( \nabla \cdot \bm{q}_i - Q_i \right) \right],
\end{equation}
\begin{equation} \label{eq:xmhd_system_4}
\notag
\textcolor{blue}{\frac{\partial T_e}{\partial t} = - \bm{u}_e \cdot \nabla T_e} \textcolor{blue}{+ \left( \gamma-1 \right) \left[ -T_e \nabla \cdot \bm{u}
- \frac{1}{n} \left( \nabla \cdot \bm{q}_e - Q_e \right) \right]},
\end{equation}
\begin{equation} \label{eq:xmhd_system_5}
\notag
\frac{\partial \bm{B}}{\partial t} = 
- \nabla \times \bm{E},
\end{equation}
with terms which are different in the single and two-temperature models colored in blue. The model is completed with the following closures:
\begin{equation} \label{eq:xmhd_system_6}
\notag
\mu_0 \bm{J} = \nabla \times \bm{B},
\end{equation}
\begin{equation}
\notag 
\bm{E} = -\bm{u} \times \bm{B} + \eta \bm{J} \\+ \frac{1}{ne} \left( \bm{J} \times \bm{B} -  \textcolor{blue}{\nabla nT_e} \right) + f_{m_e}\frac{m_e}{n e^2}
\frac{\partial \bm{J}}{\partial t},
\end{equation}
\begin{equation} \label{eq:xmhd_supp_1}
\notag
\bm{q}_s = - n \left[ \chi_{\parallel,s} \hat{\bm{b}} \hat{\bm{b}} + \chi_{\perp,s} \left( \bm{I} - \hat{\bm{b}} \hat{\bm{b}} \right) \right] \cdot \nabla T_s,
\end{equation}
\begin{equation} \label{eq:xmhd_supp_2}
\notag
Q_i = -\left( \nabla \bm{u} \right)^T : \bm{\Pi} - \textcolor{blue}{Q_\text{coll}},
\end{equation}
\begin{equation} \label{eq:xmhd_supp_3}
\notag
Q_e = \eta \bm{J}^2 + \textcolor{blue}{Q_\text{coll}},
\end{equation}
\begin{equation} \label{eq:xmhd_supp_4}
\notag
\textcolor{blue}{Q_\text{coll} = 2\times 10^{-14}\frac{n}{T_e^\frac{3}{2}}(T_i-T_e)},
\end{equation}
\begin{equation} \label{eq:xmhd_supp_6}
\notag
\bm{\Pi} = -\nu\bm{W} = -\nu( \nabla \bm{u} + \left( \nabla \bm{u} \right)^{T} - \frac{2}{3} \bm{I} \nabla \cdot \bm{u}),
\end{equation}
\begin{equation} \label{eq:xmhd_supp_7}
\notag
\bm{u}_e = \bm{u}-\frac{\bm{J}}{ne}.
\end{equation}
Here $m_i$ is the ion mass, $m_e$ is the electron mass, $e$ is the charge of an electron, $\gamma = 5$/$3$ is the adiabatic index, $\hat{\bm{b}}$ is a unit vector in the direction of the magnetic field, and $s$ indexes the fluid species, $s=i$ or $s=e$. 
An artificial particle diffusivity $D = 250$ $\text{m}^2$/$\text{s}$ is used to prevent numerical issues with small scale density oscillations. Anisotropic Braginskii thermal conduction~\cite{braginskii1965reviews,o2012simulation} $\chi_{\perp,s}$ and $\chi_{\parallel,s}$, Spitzer-like resistivity~\cite{spitzer2006physics} $\eta = \eta_0$/$T_e^{\frac{3}{2}}$, and constant and isotropic viscosity $\nu = 550$ $\text{m}^2$/$\text{s}$ are assumed. The electrons are approximated in this model to receive the entirety of the ohmic heating, while only ions receive viscous heating. These choices are well-motivated because the ratio of electron to ion ohmic heat and the ratio of ion to electron viscous heat are both approximately $(m_i$/$m_e)^{\frac{1}{2}}$. The heat exchange from ion-electron collisions, $Q_\text{coll}$, is an approximation obtained from the latest NRL plasma formulary~\cite{huba2006nrl}. 

In both numerical codes, there is an enhancement factor for the electron inertia term in Ohm's law to artificially damp Whistler waves of high spatio-temporal frequencies. This tends to reduce numerical stiffness in the magnetic field evolution. Typically
$f_{m_e} = 36.72$ and this value was found in previous work to be well-converged for simulations similar to those presented here~\cite{akcay2013extended,morgan2018finite}. 
These and other relevant parameters are summarized in Table \ref{tab:parameters}.

\begin{table}[]
    \centering
    \begin{tabular}{|c|c|}
    \hline
         \textbf{Parameter} & \textbf{Value [Units]}  \\
    \hline
         Injector Flux & 0.5 [mWb] \\
    \hline
         Injector Current & 8 [kA] \\
    \hline
         Wall Density & 0.75 [$10^{19}$ $\text{m}^{-3}$] \\
    \hline
         Wall Temperature & 3 [eV] \\
    \hline
         $\nu$ & 550 [$\text{m}^2$/$\text{s}$] \\
    \hline
         $D$ & 250 [$\text{m}^2$/$\text{s}$] \\
    \hline
         $\eta_0$ & $5.327\times 10^{-4}$ [$\Omega \text{m}$ $\text{eV}^{-\frac{3}{2}}$] \\
    \hline
         $m_i$/$(f_{m_e}m_e)$ & 100 \\
    \hline
    \end{tabular}
    \caption{Fixed parameters for PSI-Tet and NIMROD frequency scan simulations.}
    \label{tab:parameters}
\end{table}

\subsection{PSI-Tet}
PSI-Tet is a 3D high-order finite element code that supports multi-physics models on unstructured tetrahedral grids. The grids can be generated directly from Computer-Aided Design (CAD) models and this discretization facilitates the accurate representation of complex 3D geometries like the HIT-SI device. For this work the existing Hall-MHD physics module was modified to include the two-temperature model. The HIT-SI flux conserver is constructed from $1/2$" thick copper and the plasma-facing surface is coated with a thin insulating layer so that the injector operation remains purely inductive. The approximate experimental boundary conditions are
\begin{align}
\notag 
\bm{B}\cdot\hat{\bm{n}} = 0, && \bm{J}\cdot\hat{\bm{n}} = 0.
\end{align}
$\bm{B}$ is the magnetic field, $\bm{J}$ is the current density, and $\hat{\bm{n}}$ is a unit normal vector to the wall. This boundary condition is enabled by the unique mixed element discretization used by PSI-Tet. PSI-Tet additionally uses Dirichlet boundary conditions equal to the initial condition for velocity, temperature, and density, with the values
\begin{align}
\notag 
\bm{u} = 0, && T_e = T_i = 3 \text{ eV}, && n = 10^{19} \text{ m}^{-3},
\end{align}
on the boundary. 
An implicit Crank-Nicolson time advance is used with a maximum time step of 40 ns, determined to be suitable by convergence studies~\cite{hansen2015simulation}. Typical time steps are 40 ns for low frequency simulations and 10 ns for high frequency simulations. A uniform grid spacing of $2.8$ cm is used with third order basis functions. This produces an approximate resolution of $9$ mm, which resolves the smallest physical scale, the electron inertial scale $d_e \approx 12$ mm. 

\subsection{NIMROD}
NIMROD is a versatile extended MHD code used extensively for simulating spheromaks~\cite{hooper2008nimrod,akcay2013extended,morgan2017validation,morgan2018finite,morgan2019formation} as well as other plasma systems~\cite{king2016nimrod,kim2019shattered}. 
Instead of implementing the boundary condition $\bm{J}\cdot\hat{\bm{n}}$, a thin layer of high resistivity with $\eta_\text{wall}$/$\eta_\text{plasma} \approx 10^5$ is used to impede current flow into the wall. To avoid numerical issues from a sharp jump in $\eta$, a matching boundary layer is added to the mesh so that the variation occurs within a single cell. NIMROD is also restricted to toroidally symmetric geometries, so the HIT-SI injectors are modeled as boundary conditions on the flux conserver. 

The magnetic field boundary condition approximates the action of the helicity injectors through a combination of $B_\perp$ and $E_\parallel$ conditions. The spatial profile of $B_\perp$ is generated through a Grad-Shafranov solution of the injector geometry and is detailed in reference~\cite{akcay2013extended}. The injector voltage, which generates $E_\parallel = -\nabla V$, has the same spatial profile of $B_\perp$ and differs by a phase delay in time. 

The entire wall, including the injector openings, uses a constant and uniform Dirichlet boundary condition for temperature and density with the values in Table~\ref{tab:parameters}. A minor complication is that the two-temperature simulation at $f_\text{inj} = 14.5$ kHz was performed with $T_i=T_e=1$ eV because of numerical issues; we will later justify in Section~\ref{sec:temperatureScan} that the dynamics are fairly insensitive to this value. More importantly, the Dirichlet temperature and density boundary conditions for the injector may lead to significantly different dynamics than those observed in PSI-Tet. These boundary conditions enforce a cold, uniform plasma across the injectors. Parallel heat conduction then produces cold channels of plasma in the main volume which are linked to the injectors. This is in direct contrast to PSI-Tet simulations which observe highly dynamic plasma channels in the injectors which are much warmer than the wall temperature.

The velocity is zero at the boundary except for a constant inward flow at each injector mouth to counter-act density holes that cause numerical issues. These inwards flows are well-motivated by and comparable in magnitude to large flows seen in PSI-Tet simulations~\cite{hansen2014mhd,TomB}. The spatial profile of the normal velocity matches the absolute value of the normal magnetic field, and the peak velocity $\approx 19.5$ $\text{km}$/$\text{s}$ is in approximate agreement with the flow velocity observed on the experiment with ion doppler spectroscopy~\cite{hossack2015study}. 

The average cell size in the poloidal plane was 1.8 cm and used fourth order basis functions. Eleven Fourier modes in the toroidal direction were used, corresponding to a toroidal node spacing of approximately 8.3 cm at the mid-radius of the domain (27.5 cm). Grid resolution studies for both PSI-Tet~\cite{hansen2014mhd,hansen2015simulation} and NIMROD~\cite{akcay2013extended,morgan2018finite}
have shown convergence of results at these resolutions.. Lastly, NIMROD solves the same system of equations that PSI-Tet does, with the exception of a divergence cleaning term added to the magnetic field evolution~\cite{sovinec2004nonlinear}. 

\subsection{Power flows} 
Significant differences in ion and electron temperatures implies the possibility of important differences in the heating and loss terms in the evolved equations. In order to investigate these details, we calculate the various power and heat flows for the two-temperature PSI-Tet simulations. 

The injector voltage circuit is responsible for most of the power input experimentally, but in PSI-Tet only the current waveform of this circuit is known. More details can be found in the original works~\cite{hansen2014mhd,hansen2015simulation}, but for our purposes, the consequence is that the total injected power cannot be calculated directly in PSI-Tet. Instead, the total injector power $P_\text{inj}$ is approximated from a power balance of the total time rate of change of the total energy with the ion and electron heat fluxes to the wall through
\small
\medmuskip = 0.2mu 
\begin{equation}
\notag
    \sum_s\frac{d}{dt}\int_V \left[\frac{B^2}{2\mu_0} + \frac{1}{2}\rho u_s^2 + \frac{nT_s}{\gamma-1} \right]dV - \oint_\Omega \bm{q}_s\cdot \bm{d\Omega} = P_\text{inj}.
\end{equation}
We also track individual thermal energy flow terms
\begin{multline}
\notag
    \underbrace{\sum_s\frac{\partial}{\partial t}\int_V \frac{nT_s}{\gamma-1}dV}_\text{\clap{thermal Power~}} =   \underbrace{\sum_s\oint_\Omega \bm{q}_s\cdot \bm{d\Omega}}_\text{\clap{heat flux to wall~}}   +\underbrace{\int_V \nu (\nabla\bm{u})^T:\bm{W}dV}_\text{\clap{viscous heat~}} + \underbrace{\int_V \eta J^2 dV}_\text{\clap{ohmic heat~}}  +\underbrace{\frac{1}{\gamma-1}\int_V[(T_i+T_e)D\nabla^2n]dV}_\text{\clap{diffusive heat~}}
   \\ -\underbrace{\frac{1}{\gamma -1}\int_V[ (T_i+T_e)(\bm{u}\cdot\nabla n+\gamma n\nabla\cdot\bm{u})]dV}_\text{\clap{density advection and compressive heat~}}
    -\underbrace{\frac{1}{\gamma-1}\int_V [n(\bm{u}\cdot(\nabla T_i+\nabla T_e)) - \bm{J}\cdot\nabla T_e]dV}_\text{\clap{temperature advection~}}.
\end{multline}
\medmuskip = 4.0mu
\normalsize
Changes to the above relation from upwinding~\cite{BROOKS1982199}, applied to the density and temperature evolution equations, are included and contribute negligible heating power. Numerical heat flow introduced by the artificial diffusivity term is approximately $5\%$ of the total power flowing at $f_\text{inj} = 14.5$ kHz and approximately $3\%$ at $f_\text{inj} = 68.5$ kHz, a small but significant contribution.

\section{\label{sec:freqscan}Frequency Scan}
Driven nonlinear systems tend to exhibit resonances and reproduce harmonics or quasi-harmonics of the driving frequency~\cite{khalil2002nonlinear}; simulating HIT-SI with different injector frequencies provides understanding about these qualitative behavior changes. Experimental trends in injector frequency indicate that higher frequency operations tend to exhibit higher plasma impedance in the injectors, increased volume-averaged $\beta$, reduced chord-averaged density fluctuations from interferometry, and, to a lesser extent, larger current gain $G = I_\phi$/$I_\text{inj}$ . 
Previous work\cite{morgan2017validation,morgan2018finite,TomB} with the single-temperature models in NIMROD and PSI-Tet has indicated qualitative agreement with all of these trends when the injector frequency is increased from $14.5$ kHz to $ 68.5$ kHz; the two-temperature model facilitates further exploration and understanding of these qualitative trends. For the remainder of the paper, all volume-averaged quantities will be denoted like $\langle\beta\rangle$, as will the chord-averaged density $\langle n\rangle$ which is used in place of the volume-averaged density throughout the paper. The toroidal current and current centroid are obtained from averages over a number of surface magnetic probes and will be denoted $\bar{I}_\phi$ and $(\bar{R},\bar{Z})$, respectively.

Most of this analysis will be concerned with the plasma dynamics observed in these simulations, but qualitative comparisons with experimental trends will also be made. It is important to note that these simulations use typical parameters for a HIT-SI discharge at 68.5 kHz, i.e. at relatively low density and power. 
Furthermore, the total power injected in experimental HIT-SI satisfies $P_\text{inj} = V_\text{inj}I_\text{inj} = Z_\text{inj}I_\text{inj}^2 \propto f_\text{inj}I_\text{inj}^2$ where $Z_\text{inj}$ is the plasma impedance in the injector~\cite{morgan2018finite,jarboe2014proof}. These simulations keep the injector current and flux waveform amplitudes fixed, so that if the same scaling exists in the simulations, the power injected increases linearly with $f_\text{inj}$. In fact, the injector power illustration in Fig.~\ref{fig:power} indicates that the simulation produces scaling with $f_\text{inj}$ that is slightly greater than linear. While the waveform amplitudes are constant, the relative phases between the waveforms are changed to match the average phases used for high performance HIT-SI experimental discharges at each frequency, and this may also affect the injector impedance scaling with injector frequency; more details on the how the phases were chosen can be found in previous work~\cite{hansen2015simulation}.

\begin{figure*}[!tp]
    \centering
    \includegraphics[width=0.96\textwidth]{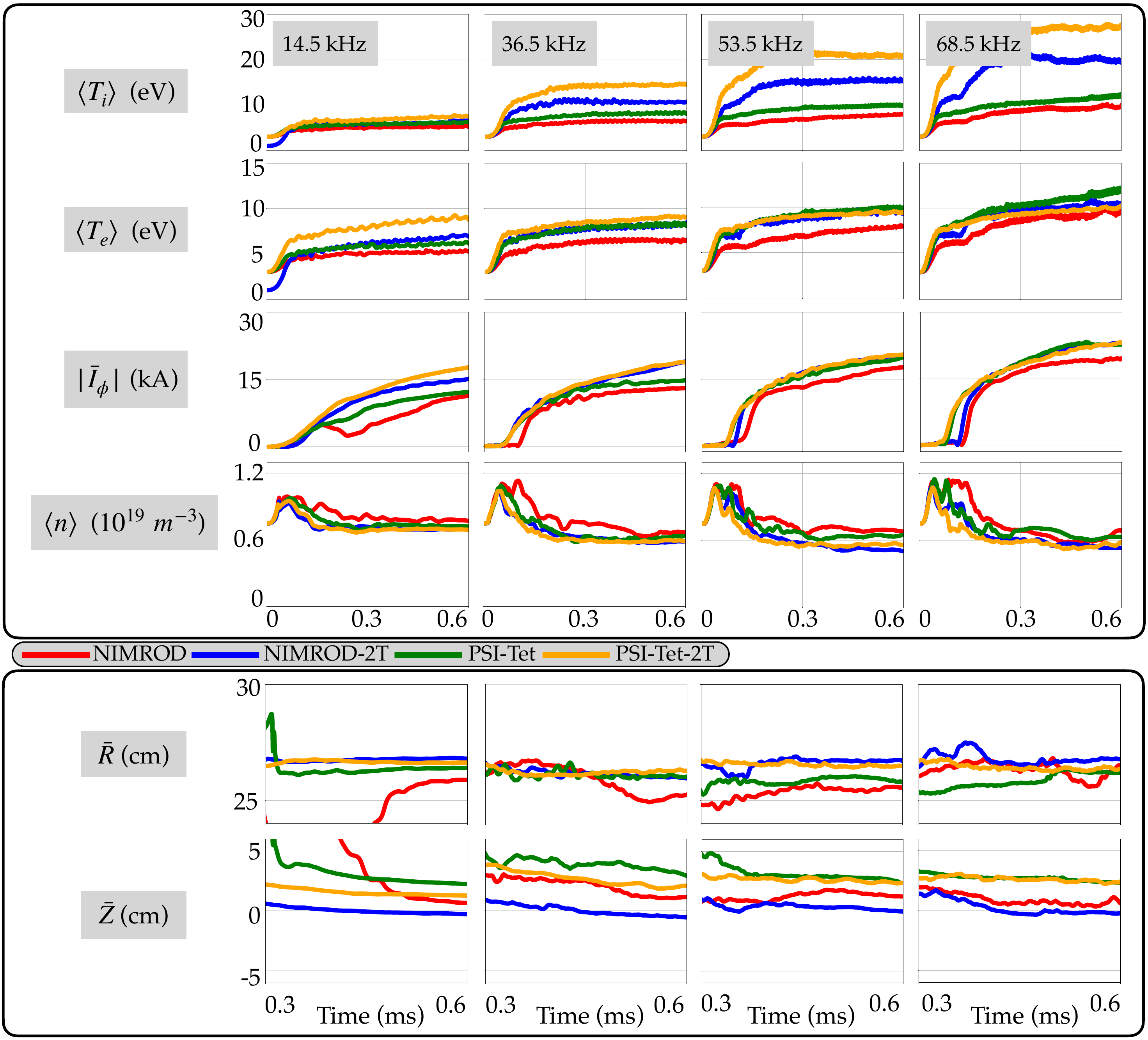}
    \caption{Summary of bulk metrics, comparing PSI-Tet and NIMROD single and two-temperature models. Density illustrations represent the chord-averaged density $\langle n \rangle$ obtained from the synthetic far-infrared interferometry diagnostic. Magnitude of the toroidal current $|\bar{I}_\phi|$ and current centroid $(\bar{R},\bar{Z})$ are calculated as averages of the four poloidal surface arrays.}
    \label{fig:bulkmetrics}
\end{figure*}

\subsection{Ion and electron temperatures}
For low density or low temperature plasmas, the slow ion-electron collision rates can result in separate temperatures for the two species on the timescales of interest. 
For the HIT-SI experiment, which sustains plasmas for 1-2 milliseconds, the approximate thermalization time $\tau_{ei} \approx 100$ $\mu$s $> \tau_\text{inj}$ for all the injector frequencies investigated here. Therefore we expect that separate ion and electron temperature evolutions are important for understanding the specific temperature profiles (affecting $\eta$ and $\beta$), as well as the flow of plasma energy and heat through the system. 

The two-temperature model shows increasing differences between $\langle T_e \rangle$ and $\langle T_i \rangle$ as the frequency increases in Fig.~\ref{fig:bulkmetrics}. At $14.5$ kHz $\langle T_e \rangle\approx \langle T_i\rangle$. However, at $68.5$ kHz $\langle T_i \rangle$ and $\langle T_e \rangle$ match closely with the experimental measurements of $T_e \approx 5-10$ eV and $T_i \approx 20-30$ eV~\cite{hossack2015study,wrobel2013}. Thomson scattering measurements on the newer HIT-SI3 device at 14.5 kHz also indicate electron temperatures of $5-10$ eV~\cite{everson2019hit}, in excellent agreement with the $\langle T_i \rangle$ calculated here. Volume-averaged electron temperature is fairly insensitive to injector frequency, while $\langle T_i \rangle$ shows a strong and approximately linear scaling with injector frequency. The mild increase for $\langle T_e \rangle$ can be partially explained through increases in the ohmic heating and electron heat flux to the wall. Collisional heating with the ions is in the tens of kilowatts and therefore plays essentially no role in the electron heating. Ohmic heating and the total electron heat flux to the wall both increase by a factor of three or four from low to high frequency, restraining volume-averaged electron temperature to a modest increase at high frequency. 

\begin{figure*}[!tp]
    \centering
    \includegraphics[width=0.97\textwidth]{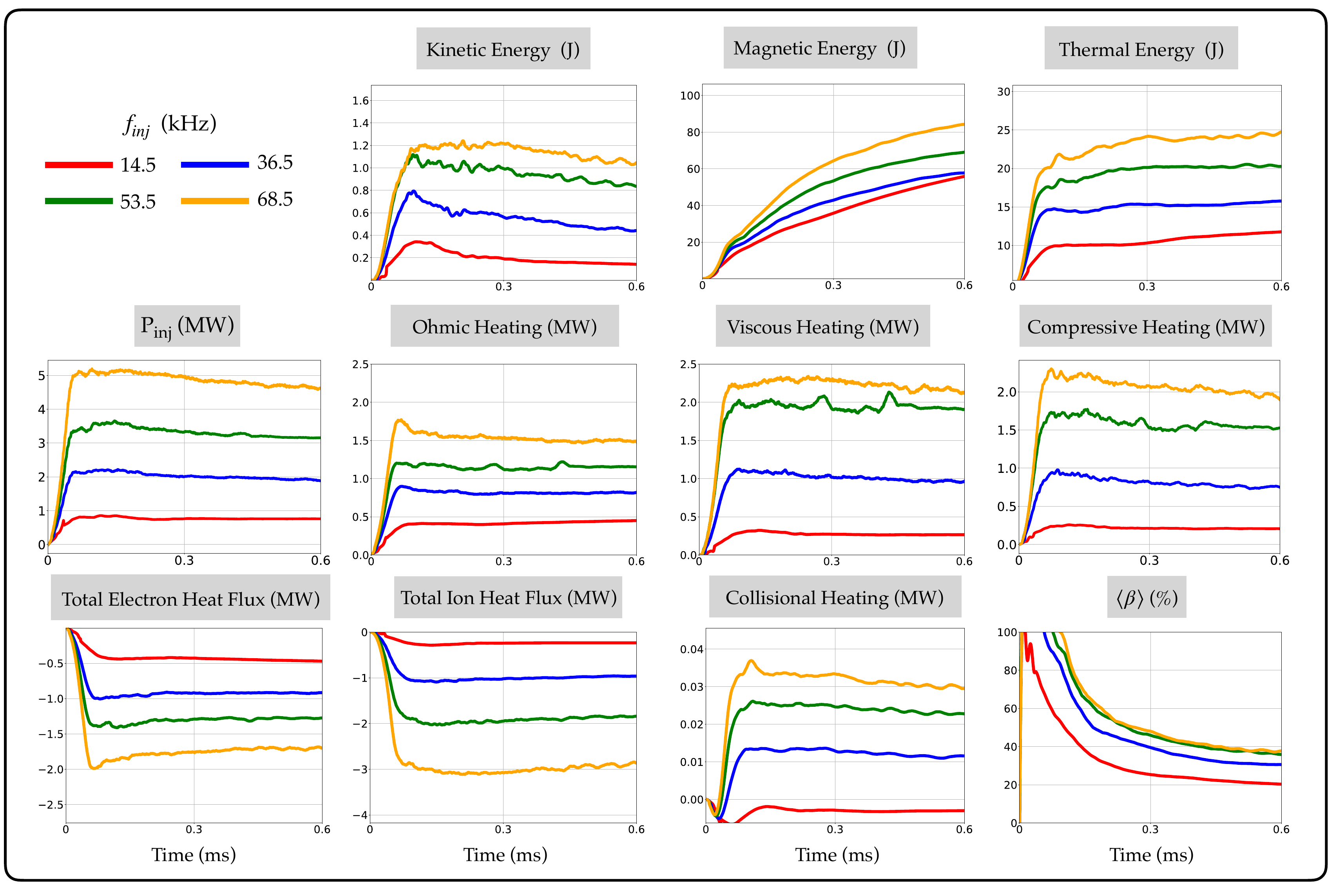}
    \caption{Summary of volume-integrated energies and power for two-temperature PSI-Tet HIT-SI simulations.}
`    \label{fig:power}
\end{figure*}

The larger volume-averaged ion temperatures can be explained from increased viscous heating. This increase is a product of more injector power at higher frequency, with a factor of approximately eight between the lowest and highest frequencies. Reconnection converts magnetic energy to ion motion, which then dissipates through viscosity, heating the ions~\cite{Ono2011,fiksel2009mass}. The change in ion motion can also be observed through the kinetic energy in Fig. \ref{fig:power}. 

The higher viscous heating at higher injector frequency can be explained physically. The frequency dependence predicted for the injector impedance is related to the energy required to do the field reversal every injector cycle. The higher the frequency, the more often the reversals happen, the more work that is expended to operate the injector, and therefore the higher the impedance. Since the injector dynamics involve significant reconnection, the viscous heating increases, leading to hotter ions. However, the large increases in the ion heat flux to the wall suggests that there is additional ion heating that offsets these losses.

A likely source of this additional heating is through the compressive heating $\propto nT_s\nabla\cdot\bm{u}$. Once the ions are warm, the compressive heat preferentially goes to the ions through the linear dependence on temperature, and Fig. \ref{fig:power} indicates that the total compressional heat $\propto n(T_e+T_i)\nabla\cdot\bm{u}$ is comparable to viscous and ohmic heating. Compression also tends to be very high in the injectors during a field reversal. The early and large (larger than ohmic for the simulations with $f_\text{inj} \geq 36.5$ kHz) viscous heating of the ions facilitates the compressive heat to preferentially heat them even further.

Lastly, single temperature models produce lower overall temperatures for both species. One possible explanation for this difference is that the heat conduction used in the single temperature case is assumed to be maximally large, since it uses the larger of the ion and electron transport in each direction. The two-temperature model develops different spatial temperature distributions that result in reduced ion temperatures near the wall, with the notable exception of inside the injectors.

To understand why the PSI-Tet two-temperature model exhibits larger $\langle T_e \rangle$ than the single-temperature model at low frequency and smaller $\langle T_e \rangle$ at high frequency, note that, at low frequency, the ohmic heating is greater than the viscous heating. The single-temperature model shares the ohmic heating with the ions, reducing the amount of heat to the electrons. At high frequency, the viscous heating is shared between the electrons and ions, so that the single-temperature model now overheats the electrons. However, the NIMROD two-temperature model has slightly larger $\langle T_e \rangle$
than the single temperature model at high frequency. This difference is possibly explained by the cold channels of plasma from the NIMROD injector boundary conditions. 

\subsection{Current centroid}
The current centroid is an important quantity for modeling magnetic equilibria and understanding the magnetic topology in HIT-SI. 
The toroidal current $\bar{I}_\phi$ and current centroid $(\bar{R},\bar{Z})$ are calculated as averages of the results from each of the four poloidal surface probe arrays in order to be consistent with the experimental analysis. 
The experimentally measured current centroid is taken as the weighted average of the poloidal field measurements from each of the four poloidal surface magnetic probe arrays~\cite{wrobel2011study}
\begin{equation}
\notag
\bar{R} = \frac{1}{4}\sum_{k=1}^4\frac{\sum_{i=1}^{16}R_{ik}B_{\theta,ik}}{\sum_{i=1}^{16}B_{\theta,ik}},
\end{equation}
\begin{equation}
\notag
\bar{Z} = \frac{1}{4}\sum_{k=1}^4\frac{\sum_{i=1}^{16}Z_{ik}B_{\theta,ik}}{\sum_{i=1}^{16}B_{\theta,ik}},
\end{equation}
where the index $k$ is summing over the four poloidal arrays and the index $i$ is summing over the 16 magnetic probes in each array. The uncertainties in the radial and vertical locations are also averaged over the four probe arrays and are quantified by the variances $\bar{\sigma}_R$ and $\bar{\sigma}_Z$.
Experimental measurements have indicated an outward radial shift and vertical symmetrization of the current centroid at high frequency in HIT-SI~\cite{victor2014sustained}. Previous work comparing single-temperature and constant-temperature simulations did not find evidence of robust changes for different injector frequencies~\cite{TomB}. The two-temperature model in PSI-Tet indicates a small but robust outward radial shift and both two-temperature models produce vertical symmetrization of the current centroid compared to single temperature models. As before, there is no consistent evidence of symmetrization or outward radial shift as the injector frequency increases.

An interesting finding is that the PSI-Tet models tend to see the vertical component of the current centroid shifted upwards $2-4$ cm from the midplane, although $\bar{\sigma}_Z$ has a similar magnitude. The change in dynamics from the single to two-temperature PSI-Tet model appears to reduce the upward shift of $\bar{Z}$ at low injector frequency, but has negligible change at high frequency. The vertical shift observations in PSI-Tet appear to be the consequence of an asymmetry between the amount of power input by the two injectors in PSI-Tet~\cite{hansen2015simulation}; injected power asymmetry is also observed experimentally. The current waveforms for the flux circuit tend to be larger for the injector on the bottom of the machine to produce the same flux in both circuits. This asymmetry depends in principle on the nonlinear plasma dynamics inside of each injector, and the stronger injector is determined by the toroidal current direction~\cite{jarboe2011recent}. In PSI-Tet simulations the bottom injector will tend to input more power, and therefore perturb the system preferentially in the upwards direction. This is directly reflected in PSI-Tet by the vertical component of the current centroid; NIMROD only produces a slight current centroid asymmetry with the single-temperature model, suggesting that capturing the injector power asymmetry correctly may require modeling the dynamics inside the injectors. In fact, the single-temperature 36.5 kHz simulation has a negative toroidal current direction and still exhibits an upward shift. 


The vertical shift of the current centroid is observed to a small extent ($\approx 1$ cm) experimentally~\cite{victor2014sustained}. The smaller experimental shift is likely because an increase in one of the injector voltage waveforms tends to occur with a corresponding decrease in the same injector current waveform, leading to approximately equal power from each injector despite the asymmetric waveforms. Dealing with this subtlety would require a more realistic model of the experimental circuit. Nonetheless, this effect provides an additional metric for experimental validation and an interesting direction for future experimental work, i.e. intentional asymmetric operation of the injectors to investigate the current centroid dependence. 

The experimentally observed shift in the radial component of the current centroid is postulated as a Shafranov shift from increased plasma pressure in the interior of the device.
In these simulations the current centroid does not consistently shift outwards with higher frequency, even with significantly larger $\langle \beta \rangle$. The volume-averaged plasma $\beta$ is pictured in Fig. \ref{fig:power} and defined here through the plasma pressure $p = n(T_i+T_e)$ as
\begin{equation}
\notag 
\langle \beta \rangle = \frac{\int_V (p-p_{wall})dV}{\int_V\frac{B^2}{2\mu_0}dV}.
\end{equation}
The lack of a shift with frequency in simulations despite an increase in $\langle \beta \rangle$ is consistent because there is no confined pressure; the field lines are all open in these simulations. 

The two-temperature PSI-Tet observations for $\langle \beta \rangle$ indicate a large increase from 14.5 kHz to 36.5 kHz (perhaps corresponding to a transition from lower to higher average $\langle \beta \rangle$ as has been demonstrated elsewhere~\cite{morgan2018finite}) and diminishing increases from 36.5 kHz to higher frequencies. These two-temperature PSI-Tet values for $\langle \beta \rangle$ are uniformly larger than those obtained with the single-temperature PSI-Tet or either of the NIMROD models. The larger values for the PSI-Tet two-temperature model may partially explain the radial outward shift of the current centroid when compared to the equivalent single-temperature model. However, there is no Shafranov shift at high frequency despite $\langle \beta \rangle$ increasing significantly with the injector frequency. The lack of an outward shift as the frequency changes is consistent with single-temperature PSI-Tet observations in previous work~\cite{TomB}. Rather than implicating confined pressure, this suggests that the current distribution changes in the two-temperature PSI-Tet simulations, which leads to the observed shift between the two-temperature and single-temperature models.

\subsection{\label{sec:impedance} Injector impedance}
\begin{figure}[!tp]
    \centering
    \begin{subfigure}{0.45\textwidth}
    \begin{overpic}[width=0.9\textwidth]{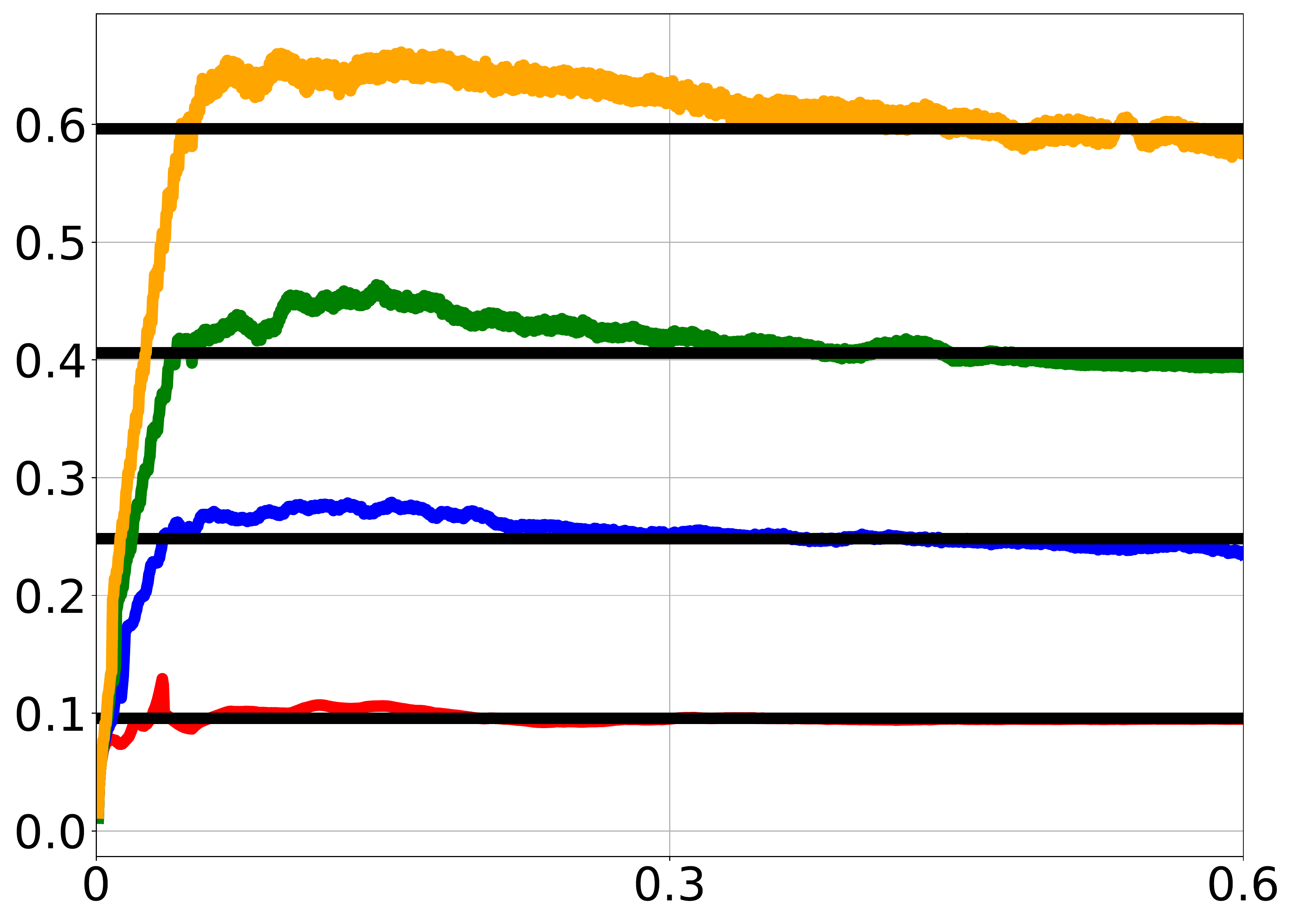}
    \put(43,-4){Time (ms)}
    \put(-10,40){$Z_\text{inj}$}
    \end{overpic}
    \end{subfigure}
    \medskip
    \vspace{.1in}
    \begin{subfigure}{0.45\textwidth}
    \begin{overpic}[width=0.90\textwidth]{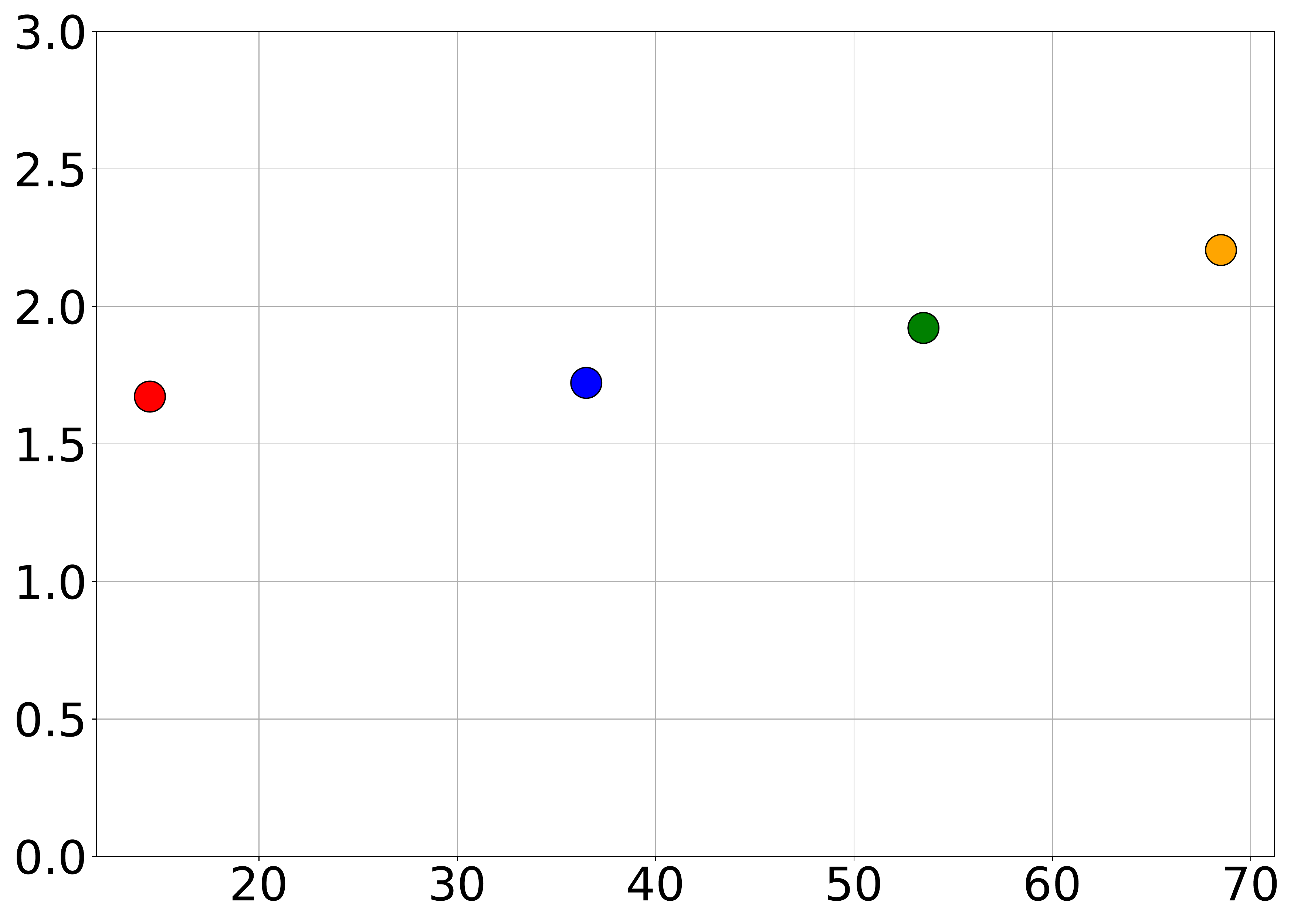}
    \put(43,-5){$f_\text{inj}$ (kHz)}
    \put(-8,40){$C$}
    \end{overpic}
    \end{subfigure}    
    \medskip
    \caption{Top: The $f_\text{inj}$ dependence captures the first order time evolution of $Z_\text{inj}(t)$. Bottom: Best fit values for $C$ corresponding to the black horizontal lines in the $Z_\text{inj}(t)$ evolution indicate averaged $C \approx 1.9$.}
    \label{fig:c2}
\end{figure}

Experimental data and imposed dynamo current drive~\cite{jarboe2012imposed,jarboe2014proof} predicts that the injector impedance approximately satisfies
\begin{align}
\notag
\label{eq:a2}
    Z_\text{inj} = C_1\mu_0R_0\left(\frac{1}{8\pi ea^3}\frac{\lambda_\text{inj}}{\lambda_\text{sph}}\frac{I_\phi}{n} + 2\pi C_2 f_\text{inj}\right). 
\end{align}
For NIMROD and PSI-Tet simulations the second term on the right-hand side is dominant~\cite{morgan2018finite}, and so we further approximate to 
\begin{align}
\notag
    Z_\text{inj} \approx 2\pi CR_0\mu_0 f_\text{inj}.
\end{align}
$R_0=0.5$ m and $a=0.25$ m are the major and minor radius of the HIT-SI device, and $C_1$, $C_2$, $C=C_1C_2$ are fitting parameters.
The $Z_\text{inj}$ scaling with the frequency is indicated in Fig. \ref{fig:c2} and indicates an average $C \approx 1.9$, in contrast to NIMROD simulations with $C \approx 0.18$. The relationship $Z_\text{inj} \propto f_\text{inj}$ appears to accurately capture the leading order injector impedance evolution, but the relationship appears to be slightly super-linear in two-temperature PSI-Tet simulations. The physical cause of this additional scaling does not appear to correlate with average $|\bm{J}|$/$n$ as suggested by IDCD. Due to the continuous evolution of quantities and their profiles in these simulations as the injector frequency changes, an alternative physical correlation could not be found.

A natural extension of this analysis is an examination of the scaling of the current gain $G = I_\phi$/$I_\text{inj}$. 
Helicity balance models~\cite{o2007fully,wrobel2013} indicate that for fixed injector waveforms and steady-state operation 
$$G \propto \sqrt{\tau_\text{L/R}Z_\text{inj}} \propto \sqrt{T_e^\frac{3}{2}f_\text{inj}},$$
where the second scaling follows from this analysis and $\tau_\text{L/R}$ is the resistive decay time. 
However, these simulations are not at steady-state operation at $t=0.6$ ms, as can be seen straightforwardly in the continued growth of the toroidal currents in Fig.~\ref{fig:bulkmetrics}. Further work should examine the scaling of gain and injector impedance at steady-state with the two-temperature models. 

\subsection{Spheromak formation}
The nonlinear relaxation event, the interval during which the plasma self-organizes into a spheromak plasma, has been analyzed extensively. However, the exact timing of this event, and the process by which the resulting toroidal current direction is determined, are not well-understood experimentally or theoretically.  
Recently, work with reduced order models using dynamic mode decomposition (DMD)~\cite{schmid2010dynamic,tu2013dynamic,brunton2019data} identified previously undiscovered large-scale magnetic structures with $n_\phi=2$ toroidal fourier structure, oscillating at the second harmonic of the injector frequency during sustainment\cite{kaptanoglu2019characterizing}. That work was performed in NIMROD on a larger and significantly hotter version of HIT-SI with a uniform and constant temperature and density Hall-MHD model~\cite{morgan2019formation}. 

\begin{figure}[!tp]
    \centering
    \begin{subfigure}{0.45\textwidth}
    \includegraphics[width=\textwidth]{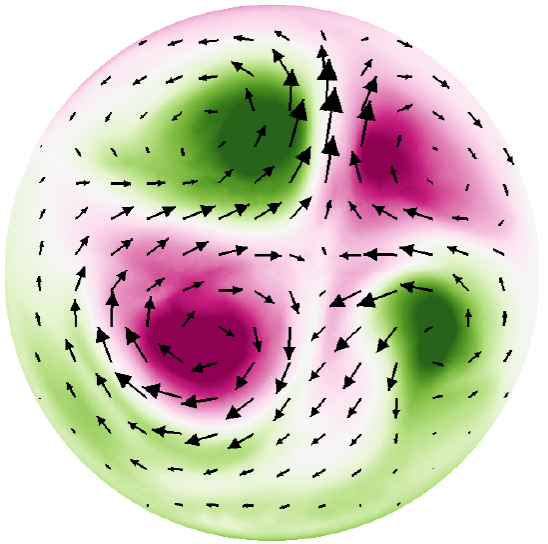}
    \caption{$f_\text{inj} = 14.5$ kHz}
    \end{subfigure}
    \begin{subfigure}{0.45\textwidth}
    \includegraphics[width=\textwidth]{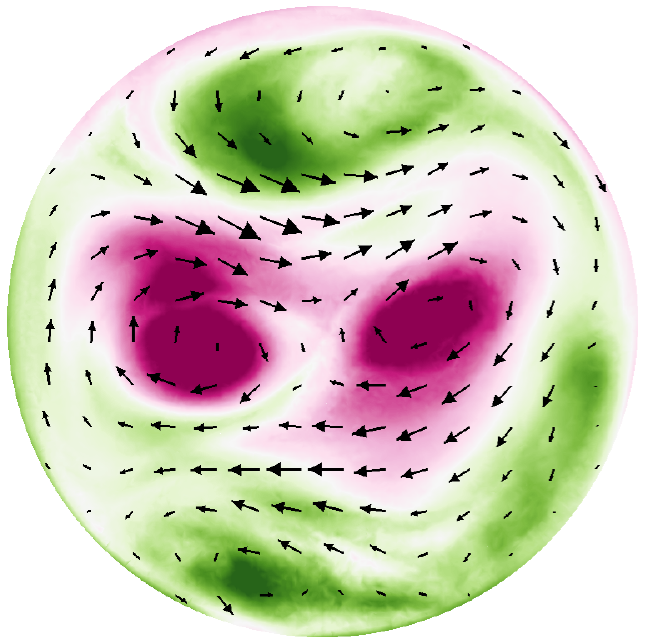}
    \caption{$f_\text{inj} = 68.5$ kHz}
    \end{subfigure}
    \caption{Contours of $B_z$ with limits $\pm 100$ Gauss and vector plots of $\bm{B}$ illustrated at time snapshots directly before spheromak formation, in the $Z=0$ midplane. The flux tubes form a $n_\phi=2$ toroidal fourier structure, and oscillate at approximately $2f_\text{inj}$. During spheromak formation, the flux tubes merge to determine the direction of the toroidal current.}
    \label{fig:flux_tubes}
\end{figure}

At high frequency, NIMROD~\cite{morgan2018finite} and PSI-Tet single and two-temperature simulations all indicate the presence of a structure of two oppositely oriented flux tubes during spheromak formation which matches the spatio-temporal dependence identified in the DMD work. Moreover, these flux tubes oscillate at approximately the second injector harmonic. This structure also briefly appears during the formation event for low frequency HIT-SI simulations using the two-temperature PSI-Tet model. The structure is visualized using $B_z$ at the $Z=0$ midplane for both low and high frequency two-temperature PSI-Tet simulations in Fig.~\ref{fig:flux_tubes}.

This observation is notable because the HIT-SI experiment tends to produce both negative and positive toroidal discharges and this parity choice appears to depend on a number of factors, including the  phases of the injector waveforms during the time of formation~\cite{morgan2018finite}. Previous studies have
found that fixed parameter discharges operating only a single injector always form and sustain spheromaks with toroidal current parity determined by the sign of the injected helicity~\cite{hossack2013reduction, Ennis_2010} but more detailed explanations for this behavior have been elusive. Moreover, HIT-SI simulations in PSI-Tet with these fixed parameters always produce negative toroidal current; equivalent simulations for the HIT-SI3 device, which has a different injector geometry, produce positive toroidal currents. This parity can often be switched in the simulations by changing the relative phases of the injector waveforms. Different parameter regimes indicate different parity preferences. 

In the simulations presented here, at the time of spheromak formation, the two closest flux tubes merge in the center, while the other two spread out and merge along the edge. Which flux tubes merge in the center determines the direction of the toroidal current. All of these observations are consistent with the interpretation of each injector driving a flux tube pair, with parity fixed by the sign of helicity. During operations with both injectors, a quasi-random process then selects which flux tube pair merges and determines the sign of toroidal current. This quasi-random process likely depends on a number of nonlinearities, as different parameter spaces indicate different timing and sign of the toroidal current. Future work could investigate how these formation structures change with different injector phasing. 

\section{\label{sec:paramscan}PSI-Tet Parameter scans}
Investigating the parameter space of the magnetohydrodynamic models presented here is essential; fixed numerical terms require convergence studies to understand their impact on simulations and physical parameters should be scanned in order to understand the possible range of results from the experimental uncertainty in the measured values. 

With the two-temperature PSI-Tet model, a number of parameter scans were performed to investigate the plasma dependence on these values. The wall temperature, wall density, and artificial diffusivity scans are performed only for $f_\text{inj} = 14.5$ kHz. Previous HIT-SI simulations have indicated only small changes from the value of the viscosity and choice of isotropic or anisotropic viscosity~\cite{morgan2017validation}. However, scans performed in NIMROD may not see dependencies on the injector geometry and associated field reversals. Future work in PSI-Tet should explore this possibility. 
\subsection{\label{sec:temperatureScan}Wall temperature}
The exact temperature boundary condition is unknown on the HIT-SI experiment. More sophisticated first-principles modeling of the temperature boundary condition would necessitate the evolution of a neutral fluid, as plasma-wall interactions involve recombination and other atomic processes which may strongly alter the density and temperature near the boundaries. However, scanning the wall temperature provides understanding about how the internal plasma dynamics are affected and provides a sense of which value best validates with experimental observations.

\begin{figure*}[!tp]
    \centering
    \includegraphics[width=0.95\textwidth]{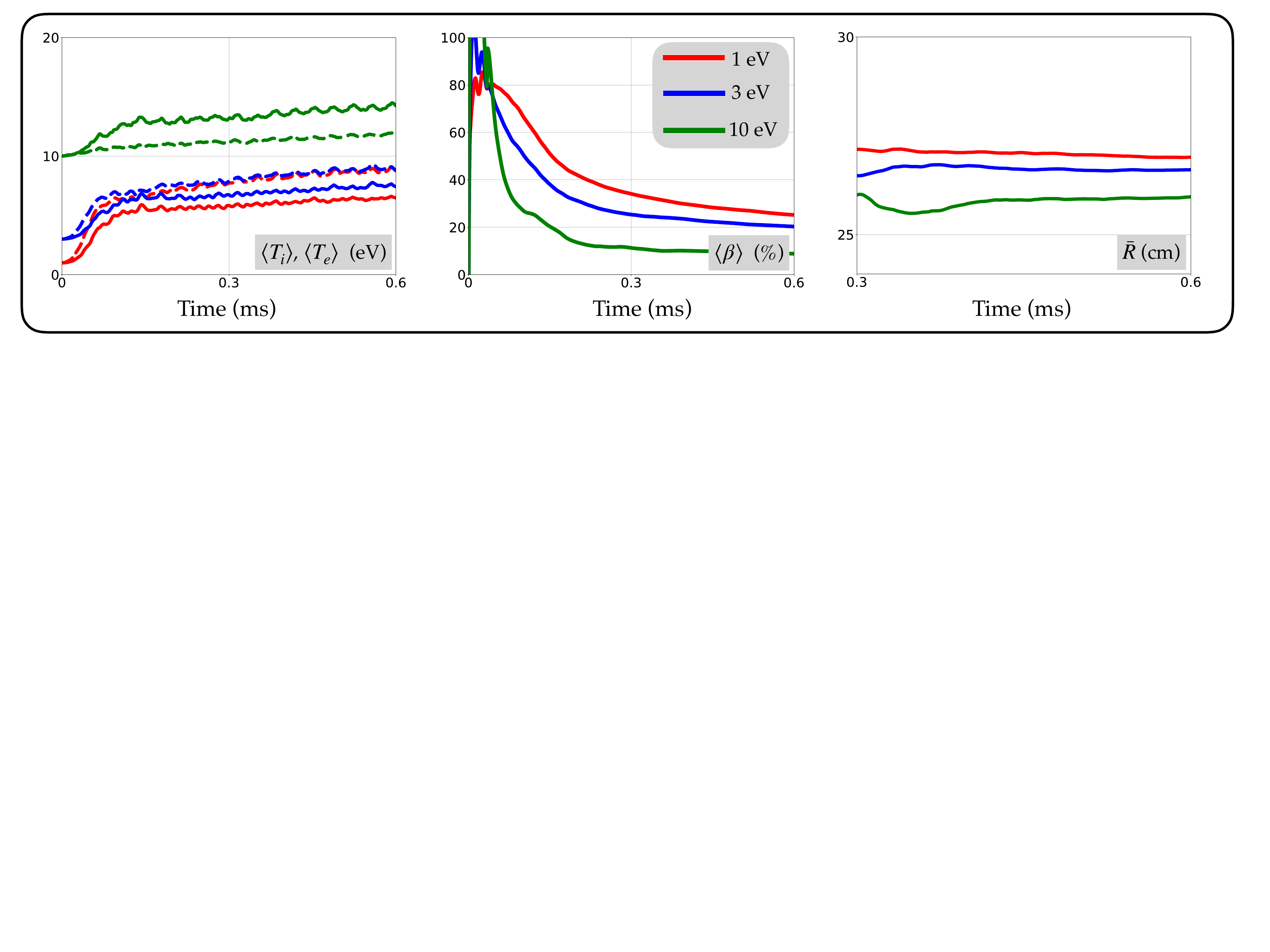}
\caption{Time evolution of important quantities for the wall temperature scan with $\langle T_i \rangle$ and $\langle T_e \rangle$ in solid and dashed lines, respectively.}
    \label{fig:Tscan}
\end{figure*}

Three different wall temperatures $T_i = T_e = 1$, 3, and 10 eV, were investigated. The only significant changes are shown in Fig.~\ref{fig:Tscan} and indicate that the primary change is a decrease in $\langle \beta \rangle$ as the wall temperature is increased, leading to a large inward shift of the current centroid. This shift is inwards because the pressure in the volume interior appears not to scale proportionally with the wall temperature. This is consistent with the interior $T_e$ not scaling proportionally with the wall temperature, while $\langle T_e \rangle$ increases substantially from the increased $T_e$ near the wall. 

\subsection{\label{sec:densityScan}Wall density}
To investigate the density dependence in simulations, wall densities $n = 10^{19}$ m$^{-3}$, $2.58\times 10^{19}$ m$^{-3}$, and $5.15\times 10^{19}$ m$^{-3}$ are scanned at $f_\text{inj} = 14.5$ kHz. To better approximate experimental conditions, injector current and flux waveform amplitudes are increased by a factor of 2.6 as compared to the other 14.5 kHz simulations shown in previous sections, leading to higher input powers of $4-8$ MW. This is comparable with the low end of experimental discharges at low frequency, which typically input $5-15$ MW of injector power.

\begin{figure}[!tp]
    \centering
    \begin{subfigure}{0.48\textwidth}
    \begin{overpic}[width=0.97\linewidth]{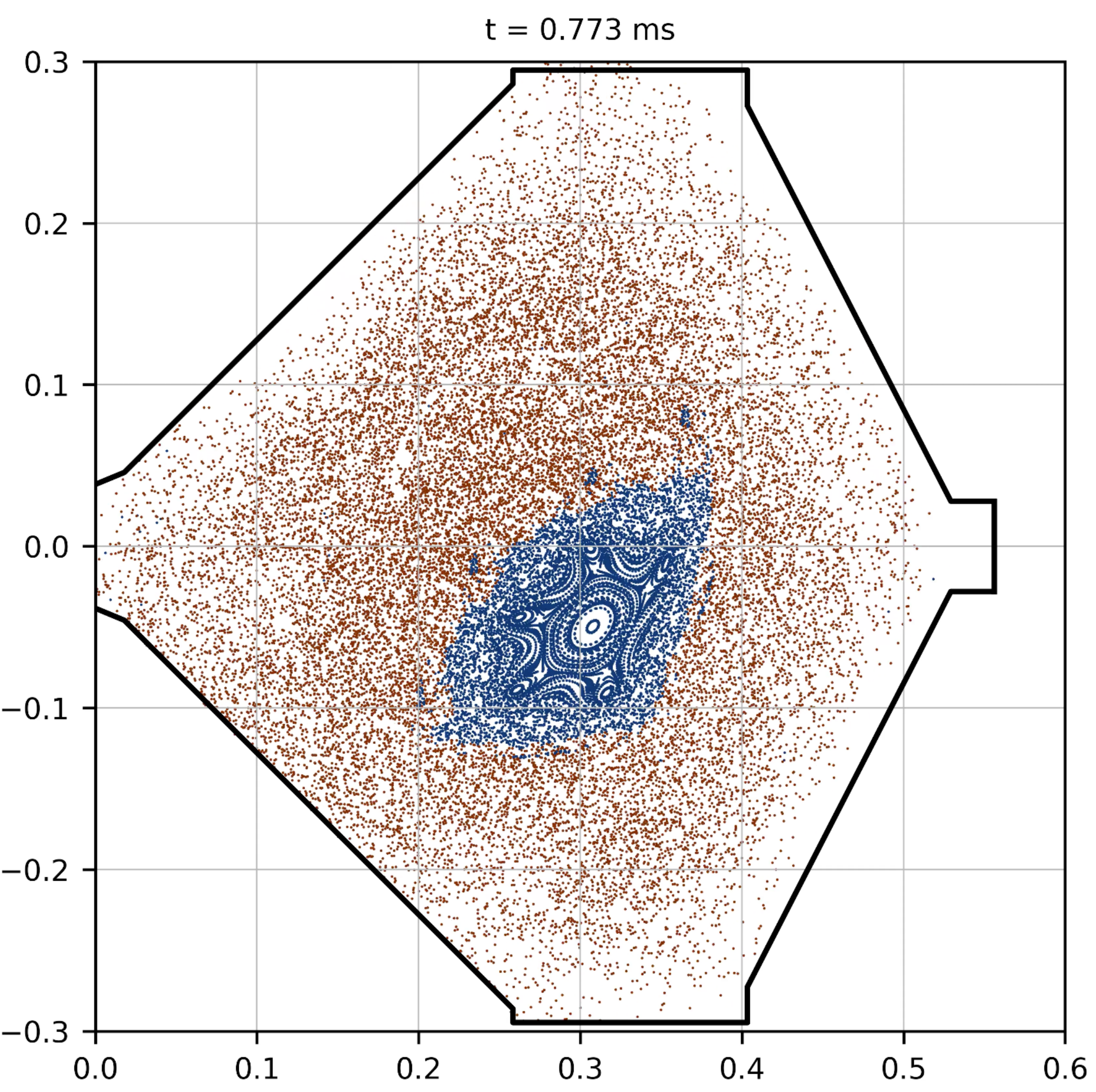}
    \put(50,-4){R (m)}
    \put(-6,55){Z (m)}
    \end{overpic}
    \vspace{.2in}
    \caption{Poincare plot of $\bm{B}$.}
    \end{subfigure}
    \begin{subfigure}{0.48\textwidth}
    \begin{overpic}[width=0.97\linewidth]{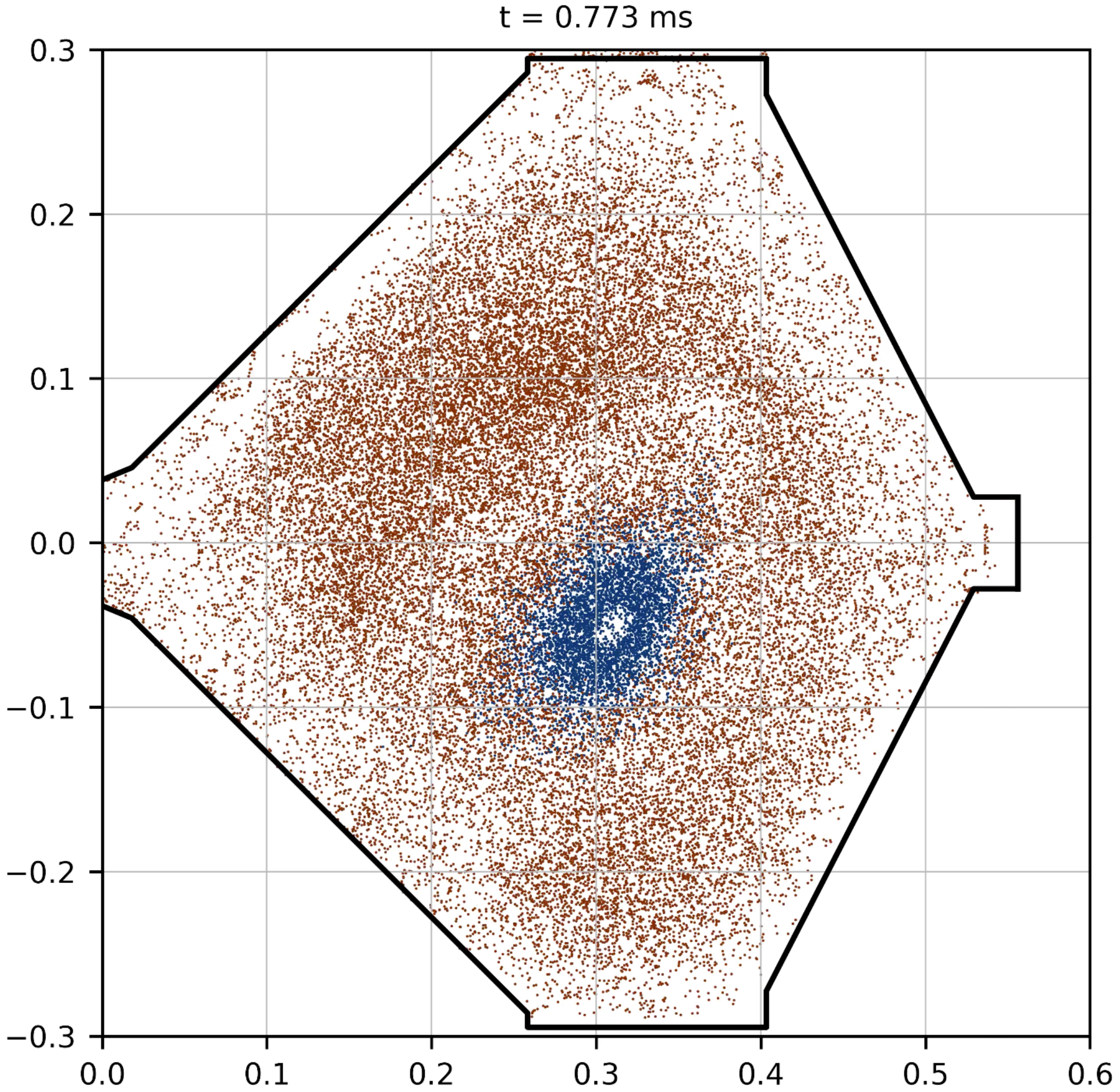}
    \put(50,-4){R (m)}
    \put(-6,55){Z (m)}
    \end{overpic}
    \vspace{.2in}
    \caption{Poincare plot of $\bm{J}$.}
    \end{subfigure}
    \caption{Closed flux surfaces with lifetimes $50-100$ $\mu$s in a low density, high power HIT-SI simulation. Points in blue approximately indicate the closed field lines.}
    \label{fig:poincare}
\end{figure}

\begin{figure*}[!tp]
    \centering
    \includegraphics[width=0.95\textwidth]{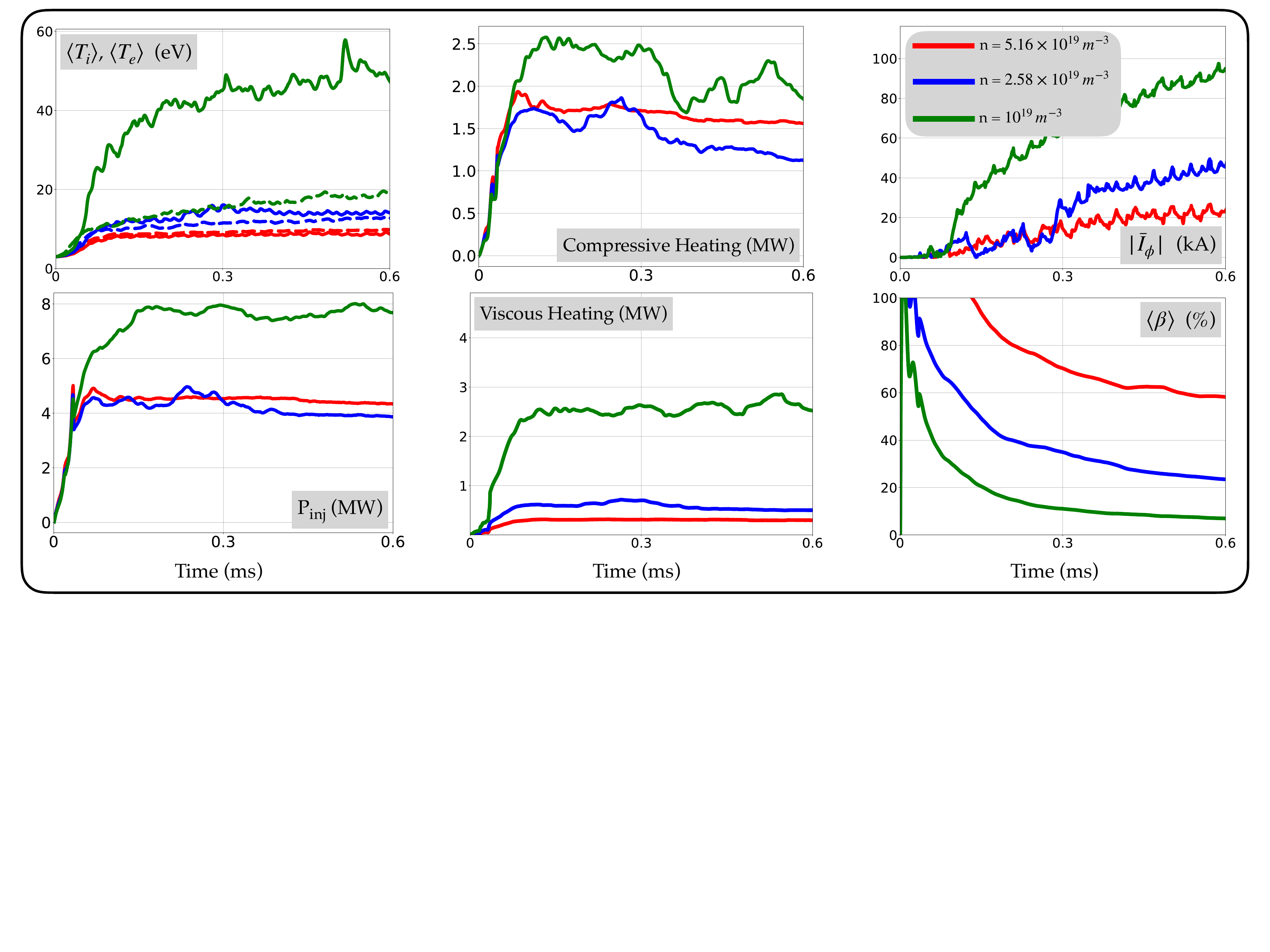}
    \caption{Time evolution of important quantities for the wall density scan with $\langle T_i \rangle$ and $\langle T_e \rangle$ in solid and dashed lines, respectively.}
    \label{fig:Nscan}
\end{figure*}

Fig. \ref{fig:Nscan} indicates that the toroidal current is strongly dependent on the density. At low density, $\langle T_i \rangle$ exceeds 50 eV, and toroidal current is approaching 100 kA and growing. Despite large gains in the temperature, $\langle \beta \rangle$ decreases, indicating the magnetic field strength has increased significantly faster than the pressure. The lowest density simulation $\langle \beta \rangle$ falls to near the Mercier limit for the HIT-SI flux conserver~\cite{mercier1963representation,mayo1988numerical}. The large magnetic field strength, the spheromak gain of $G \approx 5$, and increased fluctuations on the average temperature evolution all indicate possible closed flux activity. Poincar\'e plot movies in Fig.~\ref{fig:poincare} for $\bm{B}$ and $\bm{
J}$ confirm the existence of regions of closed flux lasting $50-100$ $\mu$s, or $1-2$ injector periods. These movies illuminate considerable variation in the closed flux surfaces. These surfaces vary from symmetric states to highly asymmetric states exhibiting complex magnetic island structures. 

The large increase in viscous heating (and therefore ion temperature) at low density can be mostly accounted for by the corresponding large increase in injector power. However, the simulation with $n=2.58\times 10^{19}$ $\text{m}^{-3}$ actually produces a slight decrease in the injector power compared to when $n=5.16\times 10^{19}$ $\text{m}^{-3}$. A similar trend is seen in the compressive heating, although it exhibits a complicated temporal dependence not observed in any other of the simulations in this work. Note that the simulation with $n=2.58\times 10^{19}$ $\text{m}^{-3}$ also indicates some interesting behavior in the compressive heat evolution, but this heat flattens out after a significant toroidal current is formed at $t\approx 0.3$ ms; the lowest density simulation, in contrast, indicates large changes in the compressive heating behavior long after it has formed a large toroidal current. All of these observations suggest a threshold at a low enough density and high enough input power where the overall performance of the device increases sharply. An experimental study at 36.5 kHz operation did not find evidence of substantial toroidal current increases with low density deuterium~\cite{hossack2013reduction}, although the total input powers were less than 4 MW for all discharges examined. The record for experimental HIT-SI gain was $G\approx 3.9$, but this high frequency discharge also had only a few MW of input power~\cite{victor2014sustained}, so the $G\approx 5$ regime has not been explored experimentally. Therefore, these observations at low density and high input power may indicate a route toward optimizing for higher performance experimental discharges. One complication in this route is that there are often large radiative losses in HIT-SI discharges, which are not modeled in the simulation. These losses may prevent the plasma from reaching the requisite viscous, ohmic, and compressive heating necessary for this higher performance regime. These findings at low density, low frequency, and high power operation merit future work on the density boundary conditions and profiles used for HIT-SI simulations. 

\subsection{Particle diffusivity}
\begin{figure*}[!tp]
    \centering
    \includegraphics[width=0.95\textwidth]{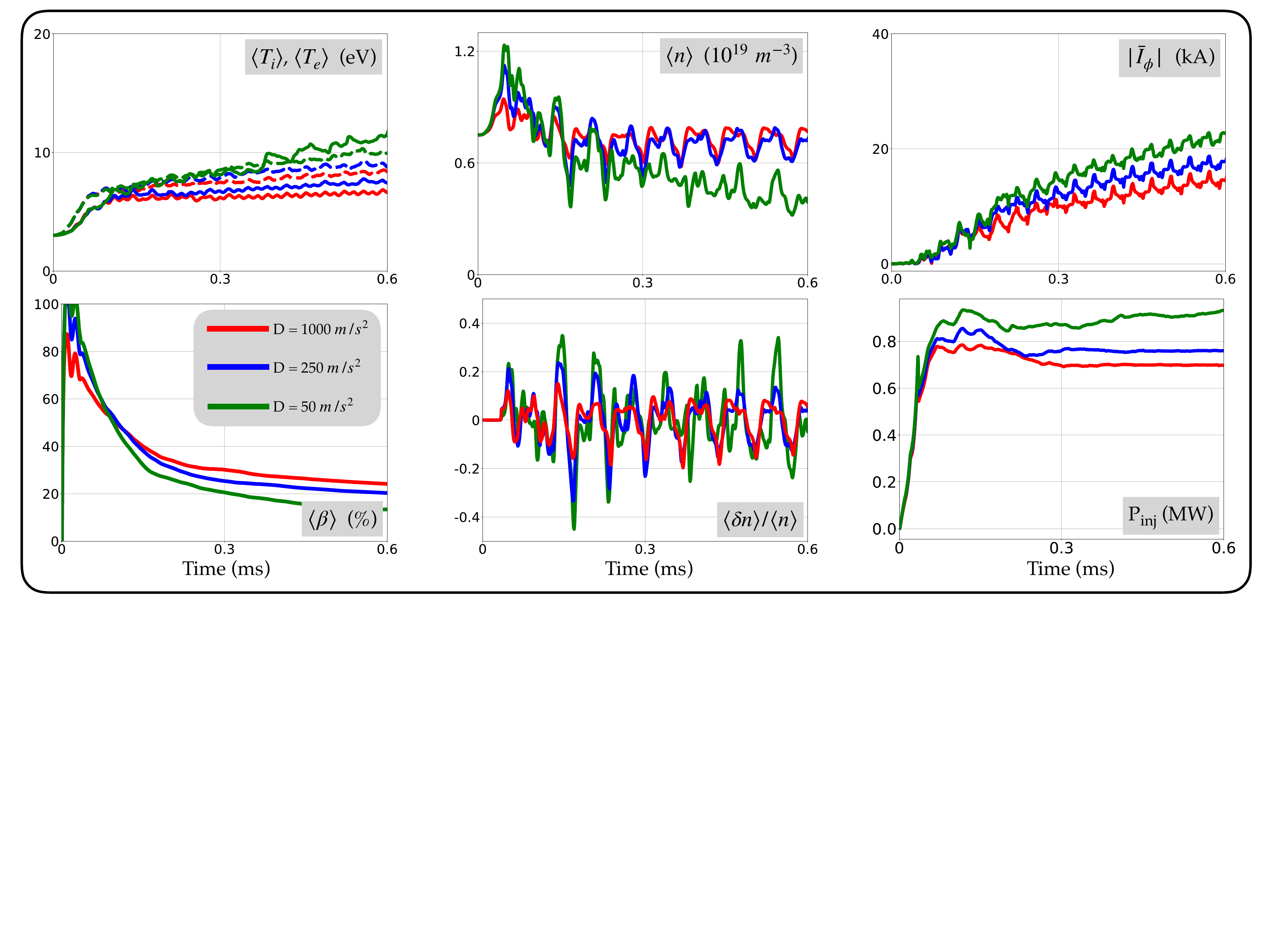}
    \caption{Time evolution of important quantities for the artificial diffusivity scan with $\langle T_i \rangle$ and $\langle T_e \rangle$ in solid and dashed lines, respectively.}
    \label{fig:Dscan}
\end{figure*}

The artificial particle diffusivity in the continuity equation is necessary for the stability of the Hall-MHD model; it is tolerable if it can be converged down to a small value where it has negligible effects on the simulation. However, previous work~\cite{morgan2018finite} only managed to reduce this value to approximately $D=250$ $\text{m}^2$/$\text{s}$. Here we report successful runs down to $D=50$ $\text{m}^2$/$\text{s}$ in PSI-Tet, which have been reproduced in NIMROD using additional hyper-diffusivity. We also compare with a run with $D=1000$ $\text{m}^2$/$\text{s}$ to understand the physical effects of this stabilizing term to regimes used in prior work~\cite{morgan2017validation}. We are unable to converge this value until the plasma dynamics are completely insensitive to it, but this comparison provides understanding about how diffusion affects the dynamics in our simulations. The quantities which indicated significant changes are summarized in Fig. \ref{fig:Dscan}.

In PSI-Tet and NIMROD this term is needed to avoid overshoot with grid-scale sharp features, generally near the wall or reconnecting regions. The two-temperature PSI-Tet simulations presented here have $\langle n \rangle$ which only mildly varies from $D = 1000$ $m^2$/$s \to 250$ $\text{m}^2$/$\text{s}$ but is significantly lower at $D = 50$ $\text{m}^2$/$\text{s}$. This reduction in chord-averaged density produces an overall increase in performance (higher toroidal current and average temperatures) consistent with the wall density scan. The volume-averaged temperatures increase significantly, and $\langle T_i \rangle > \langle T_e \rangle$ for $D = 50$ $\text{m}^2$/$\text{s}$.  Relatively unchanged thermal pressure, with significantly larger magnetic pressure at low diffusivity, leads to reduced $\langle \beta \rangle \approx 10 \%$. Despite these modest changes to the pressure balance, the current centroid shifts very little ($< 1$ cm).

Experimental interferometry exhibits large and rapid oscillations which are considerably larger than those observed in previous work or in this analysis; artificial diffusivity has been postulated as a possible explanation for this discrepancy. While the density fluctuations $\langle\delta n \rangle$ shown on the raw density signal in Fig. \ref{fig:Dscan} do not grow proportionally as the diffusivity is decreased, the relative density fluctuations $\langle\delta n \rangle $/$\langle n \rangle$ increase significantly. It is possible that further convergence of $D\to 0$ will validate better with the experimental oscillations. Another possibility is that matching the experimental density fluctuation size will require a density profile in the injectors that is a more faithful representation of the experimental fueling. Further progress will likely require careful reproduction of the experimental waveforms and possibly a direct model of the experimental circuit. 

\section{\label{sec:conclusion}Conclusions}

A two-temperature model was implemented in PSI-Tet and provides new insights into the plasma dynamics in HIT-SI simulation and experiment. Both the PSI-Tet and NIMROD two-temperature models differ from the single-temperature models through higher volume-averaged ion temperature, lower chord-averaged density, and axial symmetrization of the current centroid. Ion and electron temperatures are now in qualitative agreement with ion Doppler spectroscopy and initial Thomson scattering measurements. This new model further indicates injector impedance and volume-averaged temperatures scale approximately linearly with injector frequency. 

Parameter scans in the Dirichlet boundary condition for temperature and density, along with simulations exploring the artificial diffusivity, lead to new physical insights into the plasma dynamics in the HIT-SI device. At higher wall temperature, the interior pressure increases substantially slower than the wall pressure, leading to $\langle\beta\rangle$ decreasing and an inward shift of the current centroid. In contrast, the plasma dynamics exhibit considerable dependence on the plasma density. Performance improves with higher volume-averaged temperatures, larger toroidal current, reduced oscillations of the current centroid position, and reduced heat flux to the wall. The low density, low frequency, high power simulation indicates a sharp rise in injector power, suggesting a sudden change in the dynamics towards a higher performance regime. These simulations motivate further experimental and numerical investigation of low density parameter regimes.

Reductions in the artificial diffusivity produce significant decreases in the chord-averaged density in the device, leading to higher temperatures and toroidal currents. The relative density fluctuations increase as the chord-averaged density decreases. Further reduction may potentially lead to numerical instability near these sharp density gradients. Reduction in the artificial diffusivity in PSI-Tet may be possible by using a hyper-diffusivity term $D_h\nabla^4 n$, which can provide a larger ratio between smoothing at the grid scale and the global scale and has been successfully applied to NIMROD simulations. The observations here strongly support the claim that variations in density produce a strong impact on the dynamics.

Lastly, future work includes detailed validation with the experiment with the new two-temperature model, as has been done with previous models~\cite{hansen2015numerical}. Towards this goal, a circuit model~\cite{hooper2008nimrod} of the injectors will be implemented in PSI-Tet so that injector drive can be captured more completely by the simulation. This is expected to be important for experimental validation because the reconnection heating, propagated through the viscous heating, depends strongly on the phase between the injector waveforms during times when the injector reverses direction. The primary reversal phase was found in previous work~\cite{hansen2015simulation} to last only a few $\mu s$, consistent with the Sweet-Parker~\cite{sweet1958electromagnetic,parker1957sweet} reconnection timescale. Further physical understanding and improved validation with the experiment could also be obtained from an implementation of a more realistic closure for heat transport, along with anisotropic viscosity. 

\section{Acknowledgements}
The analysis performed in this work was funded by the U.S. Department of Energy, Office of Science, Office of Fusion Energy Sciences, under award numbers {DE-FG}02-96ER54361 and {DE-SC}0016256. Simulations presented here used resources of the National Energy Research Scientific Computing Center, supported by the Office of Science of the U.S. Department of Energy under contract number DE-AC02-05CH11231. This work was facilitated through the use of advanced computational, storage, and networking infrastructure provided by the Hyak supercomputer system and funded by the student technology fund (STF) at the University of Washington.

\bibliographystyle{plain}
 \begin{spacing}{.9}
 \small{
 \setlength{\bibsep}{6.5pt}    

 }
 \end{spacing}

\end{document}